\documentclass[11pt]{article}
\usepackage{latexsym}
\usepackage{epsf}   

\def\be{\begin{equation}}
\def\ee{\end{equation}}
\def\bea{\begin{eqnarray}}
\def\eea{\end{eqnarray}}
\def\ba{\begin{array}}
\def\ea{\end{array}}
\def\nn{\nonumber \\}

\newcommand{\ul}{\underline}

\renewcommand{\a}{\alpha}
\renewcommand{\b}{\beta}

\renewcommand{\d}{\delta}

\newcommand{\g}{\gamma}

\newcommand{\e}{\epsilon}

\begin{document}
\begin{flushright}
IFUM-751-FT\\
\end{flushright}
\vspace{1truecm}

\centerline{\Huge  Fermions, T-duality and effective actions }
\centerline{\Huge  for D-branes in bosonic backgrounds}
\vspace{2truecm}

\begin{center}
\renewcommand{\thefootnote}{\fnsymbol{footnote}}
 {\Large Donald~Marolf$^{1}$\footnote{marolf@physics.syr.edu},
 Luca~Martucci$^{2,3}$\footnote{luca.martucci@mi.infn.it}
 and Pedro~J.~Silva$^{2,3}$\footnote{pedro.silva@mi.infn.it}}\\
\renewcommand{\thefootnote}{\arabic{footnote}}
\setcounter{footnote}{0}
\vspace{.5truecm} {\small \it $^1$ Physics Department, Syracuse
University,\\ Syracuse, New York, 13244, United
States

\vspace*{0.5cm}

$^2$ Dipartimento di Fisica dell'Universit\`a di Milano,\\
Via Celoria 16, I-20133 Milano, Italy\\

\vspace*{0.5cm}

$^3$ INFN, Sezione di Milano,\\
Via Celoria 16,
I-20133 Milano, Italy\\
} 

\end{center}

\vspace{3truecm}

\centerline{\bf ABSTRACT}
\vspace{.5truecm}

\noindent
We find the effective action for any D-brane in a general bosonic background of supergravity. The results are explicit in component fields up to second order in the fermions and are obtained in a covariant manner. No interaction terms between fermions and the field $f=b+F$, characteristic of the bosonic actions, are considered. These are reserved for future work. In order to obtain the actions, we reduce directly from the M2-brane world-volume action to the D2-brane world-volume action. Then, by  means of T-duality, we obtain the other Dp-brane actions. The resulting Dp-brane actions can be written in a single compact and elegant expression.

\newpage
\section{Introduction}

In order to perform typical field theory calculations in the world-volume of the D-brane, one would like to a find convenient form of the D-brane action in general bosonic supergravity backgrounds. We present such a form in this work,
neglecting interactions between fermions and the characteristic field combination $f=b+F$. Up to such terms, these actions are by construction k-symmetric, supersymmetric\footnote{ In order to obtain a world-volume field theory with rigid supersymmetry one must fix both k-symmetry and static gauge for the coordinates.  The supersymmetry transformations are those operations that compensate for k-symmetry, gauge transformations and world-volume reparametrizations in order to preserve the chosen gauge conditions. The corresponding discussion will appear in a separate work \cite{mms}.} and in addition they obey the various dualities of M-theory.  For example the D3-brane action is self-dual. Because we allow for general curved bosonic backgrounds and the RR and NS fields act as coupling constants for the world-volume theory, the theory defined on the brane is a field theory in curved space-time with local coupling constants.

The expansion of the supersymmetric D-brane action in component fields is a non-trivial task, particularly if one uses the so called ``gauge completion'' technique (see \cite{wit,gc1} for some references where this method has been used). This method is based on a laborious comparison order-by-order between component supersymmetric transformations and superspace coordinate transformations. The numerous non-covariant gauge completions and the lack of an iterative method make the treatment very difficult and tedious, even at linear order.  In fact, no complete expansion  in general supergravity backgrounds has been obtained\footnote{There is an expansion up to second order \cite{wit}, but that work does not obtain the complete set of second order terms. There are also closed results for a few bosonic backgrounds like $AdS_5\times S^5$, $AdS_3\times S^7$, $AdS_7\times S^3$ and certain plane waves \cite{gc2}.} using this technique for any of the branes (including the M2-brane). The complications only increase at higher orders.

On the other hand one may use the so called ``normal coordinate expansion'' \cite{nc1}. This technique allows one to expand the superfield action to any order in the 32 fermions such that each term in the expansion is covariant. This technique has been used for the expansion of the heterotic superstring action and for the derivation of the superspace density formula. More recently, it has been used  for the expansion of the M2-brane up to quartic order \cite{gk1} and for the type IIa/b fundamental strings up to quadratic order \cite{ms}.

We are interested in the world-volume theory for various branes of M-theory. We will begin with the 11D supermembrane, for which the supersymmetric action has been expanded by Grisaru and Knutt in fermionic coordinates up to quartic order in general backgrounds of 11D supergravity.  We then dimensionally reduce the space-time to obtain the D2-brane action in  the 10D type IIA theory. In doing so, a world-volume duality is needed to recover the usual Yang Mills degrees of freedom in the D2-brane. Then, by means of T-duality, we obtain all the abelian D-brane actions.

Section 2 is a short review of the expansion of the supersymmetric M2-brane and a brief introduction to normal expansion techniques. Section 3 deals with the reduction of the second order expanded M2-brane action to obtain the corresponding D2-brane action. Section 4 is a comprehensive instruction manual for performing T-duality on such  actions. Section 5 gives the final form of the Dp-brane actions up to second order in the fermion expansion. Appendix A fixes our conventions for spinors and the gamma matrix algebra, appendix B gives the 11D supergravity constraints, the zero order component fields and the 10D supergravity conventions while appendix C gives some useful formulas for T-duality.


\section{Normal coordinate expansion for the M2-brane}

In a superspace formalism, the supercoordinates $z^M$ decompose into bosonic coordinates $x^m$ and fermionic coordinates $\theta^\mu$. Here we also introduce a similar decomposition for tangent space vectors $y^A$, with $A=(a,\alpha)$:
\bea
&&z^M=(x^m,\theta^\mu), \nn
&&y^A=(y^a,y^\alpha).
\eea
The normal coordinate expansion is a method based on the definition of normal
coordinates in a neighborhood of a given point $z^M$ of superspace.  
The idea is to parameterize the neighboring points by the tangent vectors
along the geodesics joining these points to the origin. Denoting the coordinates at
neighboring points by $Z^M$ and the tangent vectors by $y^A$, we have
\be
Z^M=z^M+\Sigma^M(y)\;,
\ee
where the explicit form of $\Sigma^M(y)$ is found iteratively by solving the
geodesic equation. Tensors at the point $Z^M$ may be compared with those at $z^M$ by parallel transport.  In this
sense, the change in a general tensor under an infinitesimal displacement $y^A$ is
\be
\delta T = y^A\nabla_A T\; .
\label{exp1}
\ee
Finite transport is obtained by iteration. In this way we may consider the corresponding expansion  in the operator $\delta$ for any tensor in superspace. For example,
consider the vielbein $E_M^A$
\be
E_M^{\;\;\;A}(Z)=E_M^{\;\;\;A}(z)+ \delta E_M^{\;\;\;A}(z)+ {1\over2}\delta^2E_M^{\;\;\;A}(z)+\ldots
\ee
In particular one finds the following fundamental relations by means of which one can expand any tensor iteratively to any order (see for example \cite{gk1}),
\bea
&&\delta y=0\;, \nn
&&\delta E^A= \nabla y^A + y^CE^BT_{BC}^{\;\;\;\;A}\;, \nn
&&\delta \nabla y^A = -y^BE^Cy^DR_{DCB}^{\;\;\;\;\;\;\;\;A}\;,
\eea
where $T$ is the torsion and $R$ the Riemann tensor.

In our case, we are interested in an expansion up to second order around a bosonic background of 11D supergravity. We therefore set $z^M=(x^m,0)$ and $y^A=(0,y^\alpha)$. By means of the 11D superconstraints (see appendix B), one finds the following formulas:
\bea
&&\delta E^a =0, \nn
&&\delta E^\alpha = Dy^\alpha +y^\beta e^b T_{b\beta}^{\;\;\;\alpha}\;, \nn
&&\delta^2 E^a =-i(y^\alpha\Gamma^a_{\;\;\alpha\beta}Dy^\beta+y^\alpha y^\beta T_{\;\beta}^{\;\;\;\gamma}\Gamma^a_{\;\;\gamma\alpha})\;, \nn
&&\delta^2 E^\alpha = 0\;.
\label{exp2}
\eea
At this point, in order to obtain the expanded M2-brane action in component fields, we simply take the M2-brane supersymmetric action, perform substitutions using equations (\ref{exp1},\ref{exp2}) discard terms above second order in Fermions.

The supermembrane action is,
\be
S=-T_{M2}\int{d^3\xi\sqrt{\det(-\textbf{G})}}-
{T_{M2}\over 6}\int{d^3\xi\varepsilon^{ijk}\textbf{A}_{kji}},
\ee
where $i=(0,1,2)$, $T_{M2}=(4\pi^2 l_p^{3})^{-1}$, $l_p$ is the 11D Plank length, ($\textbf{G}$,$\textbf{A}$) are the pull-back to the supermembrane of the metric and 3-form super-fields of N=1 11D supergravity.

The  normal coordinate expansion to second order in fermions yields
\bea
&&S=S^{(0)}+S^{(2)},\label{mt}\nn
&&S^{(0)}=-T_{M2}\int{d^3\xi\sqrt{-\det(G)}}-
{T_{M2}\over 6}\int{d^3\xi\varepsilon^{ijk}A_{kji}},\label{m0}\nn
&&S^{(2)}={iT_{M2}\over 2}\int d^3\xi\left\{\sqrt{-\det(G)}\left[ \bar{y}\Gamma^i\nabla_iy + \bar{y}T_i\Gamma^iy\right]+\right.\nonumber\nn
&&\;\;\;\;\;\;\;\;\;\;\;\;\;\;\;\;\;\;\;\;\;\;\;\;\;\;\;\;\;\;\;\;\;\;\;\;\
\left.-\;\hbox{${1\over 2}$}\varepsilon^{ijk}\left[\bar{y}\Gamma_{ij}\nabla_ky
+\bar{y}T_i\Gamma_{jk}y\right]\right\},
\label{M2}
\eea
where ($G,A$) are the bosonic pull-back of ($\textbf{G}$,$\textbf{A}$),
$y$ is a real Majorana 32 component spinor,  $\Gamma_i$ are pull-backs of real gamma matrices, $\nabla_i$ is the pull-back of the usual spinor covariant derivative, and
\be
T_a={1\over288}(\Gamma_a^{\;\;bcde}+8\delta^b_a\Gamma^{cde}) H_{bcde}\; ,
\ee
with $H=dA$. The above action is a truncation of the 4th order expansion found in \cite{gk1}. 

The fermionic part of action $S^{(2)}$ can be rewritten in the more compact and suggestive form,
\be
S^{(2)}=T_{M2}\int d^3\xi\sqrt{-G}\left(i\bar{y}P_-\Gamma^i\hat{D}_iy\right)\;,
\label{m2}
\ee
where
\bea
&&P_-={1\over2}(1-\hbox{${1\over 3!\sqrt{-G}}$}\epsilon^{ijk}\Gamma_{ijk}), \nn
&&\hat D_i=\nabla_i-{1\over288}(\Gamma_i^{\;\;bcde}-8\delta^b_i\Gamma^{cde}) H_{bcde}.
\eea
We would like to remark that this action is by construction supersymmetric and k-symmetric up to second order\footnote{Detailed discussions on supersymmetry and k-symmetry will appear in \cite{mms}}.


\section{M2 reduction to D2}

In this section we take the world-volume action of the supermembrane given in equations (\ref{m0},\ref{m2}) and  perform a dimensional reduction to obtain the superspace action for the type IIA D2-brane in 10D up to second order in $y$. We use hatted symbols for 11D indices and un-hatted symbols for 10D indices. As in the previous section we take $a,b,c\ldots=(0,1,\ldots,9)$ as tangent space indices and $m,n,o\ldots=(0,1,\ldots,9)$ as space-time indices. We also underline number indices corresponding to tangent space directions (i.e. $\Gamma^{\underline{0}}$), leaving space-time indices unadorned. 

Thus the bosonic space-time coordinates $x^{\hat{m}}$ split into $(x^m,x^{10})$, where all the background fields are taken to be independent of $x^{10}$. As usual, one chooses a local frame for the reduction in which one has the vielbein
\bea
E_{\hat{m}}^{\;\;\hat{a}}=\left( \begin{array}{cc}
    e^{-\phi/3}e_m^{\;\;a} & -e^{2\phi/3}C_m  \\
    0 & e^{2\phi/3}
    \end{array}\right).
\eea
Here $e_m^{\;\;a}$ is the 10D vielbein in the string frame, $C^{(1)}=dx^mC_m$ is the RR 1-form potential and $\phi$ is the dilaton. The 3-form potential of 11D supergravity $A=\hbox{${1\over 3!}$}dx^{\hat m}\wedge dx^{\hat n}\wedge dx^{\hat p}A_{\hat p \hat n \hat m}$ decomposes into the RR 3-form potential $C^{(3)}=\hbox{${1\over 3!}$}dx^m\wedge dx^n\wedge dx^pC_{pnm}$ and the NS two-form $b^{(2)}=\hbox{${1\over 2!}$}dx^m\wedge dx^nb_{nm}$ in the usual form\footnote{ We use the superspace convention for differential forms i.e. $w^{(p)}=\hbox{${1\over p!}$}dx^{m_1}\wedge\cdots \wedge dx^{m_p} w_{m_p\cdots m_1}$.},
\be
A_{mnp}=-C_{mnp} \;\;,\;\;A_{10\;mn}=b_{mn}\;.
\ee
In order that the supercoordinate transformations of 10D superspace maintain the canonical form, we also use the customary rescaling of fermions
\be
y \longrightarrow e^{-{1\over6}\phi}y\; 
\ee
in the reduction from 11D to 10D. Since we work with pull-backs to the membrane of the above fields, we define
\be
p_i=\partial_ix^{10}-\partial_ix^mC_m\ .
\ee
Hence, the pull-back of the bosonic 11D metric $G$ is written as,
\be
G_{ij}=e^{-2\phi/3}g_{ij}+e^{4\phi/3}p_ip_j\;,
\ee
where $g$ is the 10D metric.

Let us now recall that the procedure to obtain the D2-brane action with its characteristic Yang-Mills field $F=dA$, is to add a term $\hbox{${1\over2}$}\epsilon^{ijk}(p_i+C_i)F_{jk}$, in which $F_{jk}$ appears as a Lagrange multiplier, to the reduced M2-brane action.  One then takes $p_i$ as an independent variable, solves for it using its equation of motion, and inserts the result back in the action. The reduction of the bosonic action $S^{(0)}$ is known in the literature; here we just quote the result from \cite{tow}:
\be
S^{(0)}_{D2}=-T_{D2}\int{d^3\xi e^{-\phi}\sqrt{-\det(g+f)}}+
T_{D2}\int{\left(C^{(3)}-\,C^{(1)}\wedge f\right)}\;,
\label{d20}
\ee
where $T_{D2}=(4\pi^{2}l_s^3g_s)^{-1}$ is the D2-brane tension, $l_s$ is the string length, $g_s$ is the string coupling, $(C^{(1)},C^{(3)})$ are the pull-backs of the RR potentials in type IIA supergravity, $f=F+b$, $F$ is the Yang-Mills world-volume field strength, and $b$ is the pull-back of the NS-NS antisymmetric field.

We now turn to the reduction of our action (\ref{M2}). The resulting expresion contains interaction terms between the world-volume vector $p_i$ and the fermions. These terms come from fluctuations of the membrane along the 11th dimension and along the fermionic directions $y$. However, the solution obtained above for $p_i$ is of the form $p_i\propto\epsilon_{ijk}f^{jk}+ O(y^2)$. Thus, since we wish to neglect interactions between $f$ and Fermions, we may in fact consider $p_i$ to be of order $y^2$.

After some lengthly algebra we find that the bosonic part of the 10D type IIA D2-brane action is equal to (\ref{d20}), while the second order fermionic part is given by
\bea
S^{(2)}_{D2}=iT_{D2}\int d^3\xi \sqrt{-g}\bigg\{\;e^{-\phi}
\bar{y} P_{(-)}^{D2}\Gamma^i\nabla_iy \;-\hbox{${1\over 2}$} e^{-\phi}
\bar{y} P_{(-)}^{D2}\Gamma^m\partial_m\phi \;y\;+ \nonumber \\
-\; \hbox{${1\over 4\cdot 3!}$}e^{-\phi}\bar{y} P_{(-)}^{D2}\left(\Gamma^{mnp\ul{\varphi}}H_{mnp}-
3\Gamma^{mni\ul{\varphi}}H_{mni}\right)y - \hbox{${1\over 4}$}
\bar{y} P_{(-)}^{D2}\Gamma^{mi\ul{\varphi}}{\bf F}^{(2)}_{mi}y +\nonumber \\
+\;\hbox{${1\over 4\cdot 4!}$}\bar{y} P_{(-)}^{D2}\left(\Gamma^{mnpq}{\bf F}^{(4)}_{mnpq}-4\Gamma^{mnpi}{\bf F}^{(4)}_{mnpi}\right)y\bigg\}\;,
\label{d22}
\eea
and we have used
\bea
&{\bf F}^{(4)}=dC^{(3)}- C^{(1)}\wedge H\;\;,\;\;H=db\;\;,\;\; P_{(-)}^{D2}=\hbox{${1\over 2}$}
\left(1-\hbox{${1\over 3!\sqrt{-g}}$}\epsilon^{ijk}\Gamma_{ijk}\right)& \nonumber \\
&{\bf F}^{(2)}=dC^{(1)}\;\;,\;\;\Gamma^{\ul{\varphi}}=\Gamma^{\ul{01\ldots 9}}\;\;,
\;\; \nabla_iy=[\partial_i+\hbox{${1\over 4}$} w_{iab} \Gamma^{ab} ] y.&  \nonumber
\eea
See appendix B for the remaining supergravity definitions.

The full action is given by  (\ref{d20}) and (\ref{d22}) together.
Note that we have not fixed k-symmetry and that the Majorana fermion $y$ still has 32 real components.


\section{T-duality Rules}

In this section, we introduce the notation and rules necessary to perform the T-duality transformation in the fermionic part of the Dp-brane action. To do so, it is convenient to rewrite the corresponding world-volume theories in terms of combinations of the 10D supergravity fields that transform covariantly under T-duality. We will also need some basic rules to deal with the pullback on the branes of the supergravity fields.

We will perform the T-duality using the Hassan formalism\footnote{In \cite{ms} this formalism was shown to be consistent with the T-duality relation between type IIa and type IIb superstrings.} \cite{has}.
In this approach, the T-duality transformations are strongly related to the supersymmetry transformations of the gra\-vitino ($\delta \psi_m\sim D_m \epsilon$ ) and the dilatino ($\delta\lambda\sim \Delta \epsilon$). These supersymmetry transformations involve the following operators acting on 10D Majorana spinors in type IIA  supergravity (see appendix B): 
\begin{eqnarray}
D_m &=& D^{(0)}_m+W_m \cr
\Delta &=& \Delta^{(1)}+\Delta^{(2)}\ ,
\end{eqnarray}
where
\begin{eqnarray}
D^{(0)}_m &=& \partial_m +\frac{1}{4} \omega_{mab}\Gamma^{ab}+\frac{1}{4\cdot 2!}H_{mab}\Gamma^{ab}\Gamma^{\ul{\varphi}} \cr
W_m &=& \frac18 e^\phi \left( \frac{1}{2!} {\bf F}^{(2)}_{ab}\Gamma^{ab}\Gamma_m\Gamma^{\ul{\varphi}}+
\frac{1}{4!}{\bf F}^{(4)}_{abcd}\Gamma^{abcd}\Gamma_m\right)\cr
\Delta^{(1)} &=& \frac12 \left( \Gamma^m \partial_m\phi +\frac{1}{2\cdot 3!}H_{abc}\Gamma^{abc}\Gamma^{\ul{\varphi}}\right)\cr
\Delta^{(2)}&=& \frac{1}{8} e^\phi \left( \frac{3}{2!} {\bf F}^{(2)}_{ab}\Gamma^{ab}\Gamma^{\ul{\varphi}}+
\frac{1}{4!} {\bf F}^{(4)}_{abcd}\Gamma^{abcd}\right)\ .
\end{eqnarray}
It is also convenient to decompose the Majorana spinors in terms of
the Weyl spinors of type IIA and IIB.  Hence, we split our Majorana spinor
$y$ into two Majorana-Weyl (MW) spinors of opposite chirality:
\begin{eqnarray}
y=y_+ + y_-\ ,\ \Gamma^{\ul{\varphi}}\;y_\pm=\pm \; y_\pm\ .
\end{eqnarray}
When acting on MW spinors of
chirality $\pm$, the operators above take the following form,
\begin{eqnarray}
D_{(\pm)m} &=& D^{(0)}_{(\pm)m}+W_{(\pm)m}\cr
\Delta_{(\pm)} &=& \Delta^{(1)}_{(\pm)}+\Delta^{(2)}_{(\pm)}\ ,
\end{eqnarray}
with
\begin{eqnarray}
D^{(0)}_{(\pm)m} &=& \partial_m +\frac{1}{4} \omega_{mab}\Gamma^{ab}\pm \frac{1}{4\cdot 2!}H_{mab}\Gamma^{ab}\cr
W_{(\pm) m} &=& \frac18 e^\phi \left(\pm \frac{1}{2!} {\bf F}^{(2)}_{ab}\Gamma^{ab}+
\frac{1}{4!}{\bf F}^{(4)}_{abcd}\Gamma^{abcd}\right)\Gamma_m\cr
\Delta^{(1)}_{(\pm)} &=& \frac12 \left( \Gamma^m \partial_m\phi \pm\frac{1}{2\cdot 3!}H_{abc}\Gamma^{abc}\right)\cr
\Delta^{(2)}_{(\pm)}&=& \frac{1}{8} e^\phi \left( \pm\frac{3}{2!} {\bf F}^{(2)}_{ab}\Gamma^{ab}+
\frac{1}{4!} {\bf F}^{(4)}_{abcd}\Gamma^{abcd}\right)\ .
\end{eqnarray}
In the rest of this work, we will not write the subscript $(\pm)$
explicitly, as it will be determined by the chirality of
the spinor on which the operators act.

For type IIB supergravity theory, we have two MW spinors $y_{(1,2)}$
of positive chirality and the following operators acting on them
(the upper sign refers to $y_1$ while the lower one to $y_2$):
\begin{eqnarray}
\hat D^{(0)}_{(1,2)m} &=& \partial_m +\frac{1}{4} \omega_{mab}\Gamma^{ab}\pm \frac{1}{4\cdot 2!}H_{mab}\Gamma^{ab}\cr
\hat W_{(1,2) m} &=& \frac18 e^\phi \left(\mp {\bf F}^{(1)}_a\Gamma^a - \frac{1}{3!} {\bf F}^{(3)}_{abc}\Gamma^{abc}\mp
\frac{1}{4!}{\bf F}_{abcd}\Gamma^{abcd}\right)\Gamma_m\cr
\hat\Delta^{(1)}_{(1,2)} &=& \frac12 \left( \Gamma^m \partial_m\phi \pm\frac{1}{2\cdot 3!}H_{abc}\Gamma^{abc}\right)\cr
\hat\Delta^{(2)}_{(1,2)}&=& \frac{1}{2} e^\phi \left( \pm  {\bf F}^{(1)}_{a}\Gamma^{a}+
\frac{1}{2\cdot 3!} {\bf F}^{(3)}_{abc}\Gamma^{abc}\right)\ .
\end{eqnarray}
As for the type IIA operators, we will suppress the subscript
$(1,2)$.  This subscript is determined by the spinor on which the operators act.
\subsection{T-duality rules for background fields}

We wish to apply T-duality along the 9th direction. Let us introduce
the following useful objects ( $\hat m,\hat n=0,\ldots,8$)
\begin{eqnarray}
&& \Omega=\frac{1}{\sqrt{g_{99}}}\Gamma^{\ul{\varphi}}\Gamma_9 \Rightarrow \Omega^2=-1\cr
&& E_{mn}=g_{mn}+b_{mn}\cr
&& (Q_{\pm})^m{}_n=\left(
\begin{array}{cc}
\mp g_{99} & \mp (g\mp b)_{9\hat n}\\
 0         &   {\bf 1}_9 \\
\end{array} \right)\cr
&& (Q^{-1}_{\pm})^m{}_n=\left(
\begin{array}{cc}
\mp g_{99}^{-1} & -g_{99}^{-1}(g\mp b)_{9\hat n}\\
 0         &   {\bf 1}_9 \\
\end{array} \right)\ .
\end{eqnarray}
The T-duality rules for $E_{mn}$ and $\phi$ are\footnote{Here and in the rest of this work, 
we place a tilde over the transformed fields to remove ambiguity when needed.},
\begin{eqnarray}
\tilde E _{\hat m\hat n}&=& E _{\hat m\hat n}-E _{\hat m 9}g_{99}^{-1}E _{9\hat n}\cr
\tilde E _{\hat m 9}&=& E _{\hat m 9}g_{99}^{-1}\cr
\tilde E _{9 \hat m }&=& -E _{9\hat m }g_{99}^{-1}\cr
\tilde E _{9 9 } &=& g_{99}^{-1}\cr 
\tilde \phi &=& \phi -{1\over 2}ln g_{99}\ .
\end{eqnarray}
For the transformation of the vielbein and the spinors, we will use the Hassan
conventions to avoid ambiguities\footnote{Recall that there are two possible choices $e^m_{(\pm)a}$ for the transformed vielbein, related by a Lorentz transformation $\Lambda^b{}_{a}$.}.
The transformation rules for the vielbein are
\begin{eqnarray}
\tilde e^m_a\equiv e^m_{(-)a}= (Q_{-})^m{}_n e ^n_a \Rightarrow \tilde e ^a_m\equiv e^a_{(-)m}= (Q_{-}^{-1})^n{}_m e ^a_n\ .
\end{eqnarray}
We will also need the alternative transformed vielbein
\begin{eqnarray}
 e^m_{(+)a}= (Q_{+})^m{}_n e ^n_a=\Lambda^b{}_{a} e^m_{(-)b} \Rightarrow e^a_{(+)m}=
(Q_{+}^{-1})^n{}_m e ^a_n=e^b_{(-)m}\Lambda_b{}^a\ ,
\end{eqnarray}
and the transformation rules for the RR potentials $C^{(n)}$
\bea
\tilde C^{(n)}_{9\hat m_2\cdots \hat m_n}&=&C^{(n-1)}_{9\hat m_2\cdots \hat m_n}-(n-1)g^{-1}_{99}g_{9[\hat m_2}C^{(n-1)}_{9\hat m_3\cdots \hat m_n]}\;, \nn
\tilde C^{(n)}_{\hat m_1\cdots \hat m_n}&=&C^{(n+1)}_{9\hat m_1\cdots \hat m_n}-nb_{9[\hat m_2}\tilde C^{(n)}_{9\hat m_2\cdots \hat m_n]}\;.
\label{trr}
\eea
Therefore going from IIA to IIB, we have:
\begin{eqnarray}
y_+ &=& y_1 \Rightarrow  \bar y_+ = \bar y_1\cr
y_- &=& -\Omega y_2 \Rightarrow  \bar y_- = \bar y_2 \Omega\cr
&&\cr
D^{(0)}_m y_+ &=& (Q_{+}^{-1})^n{}_m(\hat D^{(0)}_n y_1)\cr
D^{(0)}_m y_- &=& -\Omega(Q_{-}^{-1})^n{}_m(\hat D^{(0)}_n y_2)\cr
W_m y_+ &=& -\Omega(Q_{-}^{-1})^n{}_m(\hat W_n y_1)\cr
W_m y_- &=& (Q_{+}^{-1})^n{}_m(\hat W_n y_2)\cr
&&\cr
\Delta^{(1)}y_+ &=& \hat \Delta^{(1)}y_1 -g_{99}^{-1}\Gamma_9 \hat D ^{(0)}_9 y_1\cr
\Delta^{(1)}y_- &=& -\Omega(\hat \Delta^{(1)}y_2 -g_{99}^{-1}\Gamma_9 \hat D ^{(0)}_9 y_2)\cr
\Delta^{(2)}y_+ &=& -\Omega(\hat \Delta^{(2)}y_1 -g_{99}^{-1}\Gamma_9 \hat W_9 y_1)\cr
\Delta^{(2)}y_- &=& \hat \Delta^{(2)}y_2 -g_{99}^{-1}\Gamma_9 \hat W_9 y_2\ .
\end{eqnarray}
Conversely, going from IIB to IIA we have
\begin{eqnarray}
y_1 &=& y_+ \Rightarrow  \bar y_1 = \bar y_+\cr
y_2 &=& \Omega y_- \Rightarrow  \bar y_2 = -\bar y_- \Omega\cr
&&\cr
\hat D^{(0)}_m y_1 &=& (Q_{+}^{-1})^n{}_m (D^{(0)}_n y_+)\cr
\hat D^{(0)}_m y_2 &=& \Omega(Q_{-}^{-1})^n{}_m(D^{(0)}_n y_-)\cr
\hat W_m y_1 &=& \Omega(Q_{-}^{-1})^n{}_m(W_n y_+)\cr
\hat W_m y_- &=& (Q_{+}^{-1})^n{}_m(W_n y_-)\cr
&&\cr
\hat\Delta^{(1)}y_1 &=& \Delta^{(1)}y_+ -g_{99}^{-1}\Gamma_9 D ^{(0)}_9 y_+\cr
\hat\Delta^{(1)}y_2 &=& \Omega(\Delta^{(1)}y_- -g_{99}^{-1}\Gamma_9 D ^{(0)}_9 y_-)\cr
\hat\Delta^{(2)}y_1 &=& \Omega(\hat \Delta^{(2)}y_+ -g_{99}^{-1}\Gamma_9  W_9 y_+)\cr
\hat \Delta^{(2)}y_2 &=& \Delta^{(2)}y_- -g_{99}^{-1}\Gamma_9 W_9 y_-\ .
\end{eqnarray}

\subsection{T-duality rules for pulled-back quantities}
\label{td}

Once again we will need to fix some conventions. We use 
world-sheet coordinates $\xi^i$ ($i=0,\ldots,p$) on the
$Dp$-brane and $\xi^I$ ($i=0,\ldots,p,9$) on the $D(p+1)$-brane.
We also fix the static gauge $x^i(\xi)=\xi^i$ for the
$Dp$-brane and $x^I(\xi)=\xi^I$ for the $D(p+1)$-brane.
We adopt Myers' notation \cite{myers} for the pull-back
($\tilde m =p+1,\ldots,9$, $\hat m =p+1,\ldots,8$):
\begin{eqnarray}
&& P_{Dp}[M_i]=\partial_i x^m M_m =M_i +\partial_i x^{\tilde m} M_{\tilde m}\cr
&& P_{D(p+1)}[M_I]=\partial_I x^m M_m =M_I +\partial_I x^{\hat m} M_{\hat m}\ ,
\end{eqnarray}
and $P_{Dp}[g^{ij}]$ stands for the inverse of $P_{Dp}[g_{ij}]$.

As above, we consider the field $f_{ij}=P_{Dp}[B_{ij}]+F_{ij}$ to be of order $O(y^2)$ for any Dp-brane and we neglect the interaction terms of order $O(y^3)$. The T-duality transformation rules for the pulled-back fields are then straightforward but tedious to derive. Since they are not very illuminating and may result in a undesired distraction for the reader, we have relegated them to appendix C.


\section{Dp-brane actions}

The T-duality rules described in the previous section, allow us to obtain any of the type II super D-brane actions from the D2-brane action of equation (\ref{d20},\ref{d22}). To do this, we first use the well-known fact that the bosonic Dp-brane actions follow by T-duality from that of the bosonic D2-brane, i.e.
\begin{eqnarray}
\label{123action}
iT_{D2}\int{d^3\xi e^{-\phi}\sqrt{-(g+f)}}\ \rightarrow \
iT_{Dp}\int{d^{p+1}\xi e^{-\phi}\sqrt{-(g+f)}}\ ,
\end{eqnarray}
and
\bea
T_{D2}\int \Sigma_n \;C^{(n)}\;e^{-f}\ \rightarrow \ T_{Dp}\int \Sigma_n \;C^{(n)}\;e^{-f},
\eea
where $T_{Dp}=2\pi[(2\pi l_s)^{p+1}g_s]^{-1}$. Second, we rewrite the fermionic part of the action (\ref{d22}) in a much more compact, elegant, and appropriate form for the study of T-duality,
\begin{eqnarray}\label{fermionic}
S^{(2)}_{D2}=iT_{D2}\int{d^3\xi\;e^{-\phi}\sqrt{-g}\;\bar{y}\;P^{D2}_{(-)}(\Gamma^iD_i -\Delta)y}\;,
\end{eqnarray}
using the definitions introduced in the previous section.  Finally, we observe that, up to interacting terms in $f$ and $y$, under T-duality we have
\begin{eqnarray}
iT_{D2}\int{d^3\xi e^{-\phi}\sqrt{-g}\ldots}\ \rightarrow \
iT_{Dp}\int{d^{p+1}\xi e^{-\phi}\sqrt{-g}\ldots}\hspace{0.5cm} .
\end{eqnarray}
Hence, we need only study the T-duality transformation of the quadratic terms in $y$ to obtain all the fermionic parts of the Dp-brane actions. 

Let us consider the case of a T-duality transformation to the $D3$-brane; i.e. a T-duality along a direction $x^9$ transverse to the original D2-brane. To do so, we separate the fermion $y$ in two fermions of opposite chirality $y=y_+ + y_-$. Then, using the definitions introduced in the previous section, we have
\begin{eqnarray}
\bar{y}\;P^{D2}_{(-)}(\Gamma^iD_i -\Delta)y &=& \frac12\big\{ \bar{y}_+(\Gamma^iD^{(0)}_i -\Delta^{(1)})y_+
-\bar{y}_+\Gamma_{D2}(\Gamma^i W_i -\Delta^{(2)})y_+ +\cr
&& +\bar{y}_-(\Gamma^i W_i -\Delta^{(2)})y_+ -\bar{y}_-\Gamma_{D2}(\Gamma^iD^{(0)}_i -\Delta^{(1)})y_+ +\cr
&& +\bar{y}_+(\Gamma^i W_i -\Delta^{(2)})y_- - \bar{y}_+\Gamma_{D2} (\Gamma^iD^{(0)}_i -\Delta^{(1)})y_- +\cr
&& \bar{y}_-(\Gamma^iD^{(0)}_i -\Delta^{(1)})y_- -\bar{y}_-\Gamma_{D2} (\Gamma^i W_i -\Delta^{(2)}) y_-\big\}\ .   
\end{eqnarray}

Consider for example the term  $\bar{y}_+(\Gamma^iD^{(0)}_i -\Delta^{(1)})y_+$.  This may be rewritten more explicitly as follows:
\begin{eqnarray}
\label{backwards}
\bar{y}_+ P_{D2}[\tilde g^{ij}] P_{D2}[\Gamma_{(-)i}] P_{D2}[D^{(0)}_j y_+]-  \bar{y}_+ \Delta^{(1)}y_+ \ .
\end{eqnarray}
It is convenient to regard the result (\ref{backwards}) as having been obtained by applying T-duality to a D3-brane, so that our goal is to reconstruct the ``original'' type IIB term in the D3-brane action. 
Applying the rules of the previous section and appendix C, one finds that, up to interaction terms in $f$ and $y$, we have
\begin{eqnarray}
&&\bar{y}_+ P_{D2}[\tilde g^{ij}] P_{D2}[\Gamma_{(-)i}] P_{D2}[D^{(0)}_j y_+]\cr
&&= \bar{y}_1 P_{D3}[g^{ij}] P_{D3}[\Gamma_{i}] P_{D3}[\hat D^{(0)}_j y_1]-\cr
&& - \bar{y}_1 P_{D3}[g^{ij}] P_{D3}[\Gamma_{9}] P_{D3}[g_{99}]^{-1}P_{D3}[g_{9i}]P_{D3}[\hat D^{(0)}_j y_1]-\cr
&& -\bar{y}_1 P_{D3}[g^{ij}] P_{D3}[\Gamma_{i}] P_{D3}[g_{99}]^{-1}P_{D3}[g_{9j}]P_{D3}[\hat D^{(0)}_9 y_1]+\cr
&& +\bar{y}_1 P_{D3}[g^{ij}] P_{D3}[\Gamma_{9}] P_{D3}[g_{99}]^{-1}P_{D3}[g_{9i}]P_{D3}[g_{9j}]P_{D3}[\hat D^{(0)}_9 y_1]\ ,\cr
&&\cr
and &&\bar{y}_+ \Delta^{(1)}y_+ = \bar{y}_1 \hat \Delta^{(1)}y_1 - P_{D3}[\Gamma_{9}] P_{D3}[g_{99}]^{-1}[\hat D^{(0)}_9 y_1]\ .
\end{eqnarray}
Now for any matrix $M_{IJ}$ with inverse $M^{IJ}$, one has the identities
\begin{eqnarray}
M^{i9} &=& -(M_{99})^{-1}M^{ij}M_{j9} \cr
M^{99} &=&  (M_{99})^{-1}+(M_{99})^{-2}M_{9i}M^{ij}M_{j9}\ ,
\end{eqnarray}
from which we obtain
\begin{eqnarray}
&&\bar{y}_+ P_{D2}[\tilde g^{ij}] P_{D2}[\Gamma_{(-)i}] P_{D2}[D^{(0)}_j y_+]-  \bar{y}_+ \Delta^{(1)}y_+ =\cr
&&=\bar{y}_1 P_{D3}[g^{IJ}] P_{D3}[\Gamma_{I}] P_{D3}[\hat D^{(0)}_J y_1]-  \bar{y}_1 \hat \Delta^{(1)}y_1 \ .
\end{eqnarray}
All other terms of the fermionic D2-brane action transform similarly. The fermionic part of the $D3$-brane action is thus
\begin{eqnarray}
S^{(2)}_{D3}&=& \frac{i}{2} T_{D3}\int d^4\xi e^{-\phi}\sqrt{-g}\big\{ \bar{y}_1(\Gamma^I \hat D^{(0)}_I -\hat\Delta^{(1)})y_1+\cr
&&+i \bar{y}_1\Gamma_{D3}(\Gamma^I \hat W_I -\hat\Delta^{(2)})y_1 +\bar{y}_2(\Gamma^I\hat W_I -\hat\Delta^{(2)})y_1 -\cr
&&-i\bar{y}_2\Gamma_{D3}(\Gamma^I \hat D^{(0)}_I -\hat\Delta^{(1)})y_1 +\bar{y}_1(\Gamma^I \hat W_I -\hat\Delta^{(2)})y_2 +\cr
&&+ i \bar{y}_1\Gamma_{D3} (\Gamma^I\hat D^{(0)}_I -\hat\Delta^{(1)})y_2 + \bar{y}_2(\Gamma^I \hat D^{(0)}_I
 -\hat\Delta^{(1)})y_2 -\cr && -i\bar{y}_2\Gamma_{D3} (\Gamma^I \hat W_I -\hat\Delta^{(2)}) y_2\big\}\ ,
\end{eqnarray}
where we have introduced
\begin{eqnarray}
\Gamma_{D3}=\frac{-i}{4!\sqrt{-P_{D3}[g]}}\epsilon^{IJKR}P_{D3}[\Gamma_{IJKR}]\ .
\end{eqnarray}

To rewrite $S^{(2)}_{D3}$ in more compact form, we define the quantities 
\begin{eqnarray}
y &=& \left( \begin{array}{c} y_1 \\ y_2 \end{array}\right) \cr
\breve{D}_m &=& \hat D_m^{(0)}+\sigma_1\otimes \hat W_m \cr
\breve \Delta &=&  \hat \Delta^{(1)} +\sigma_1 \otimes \hat \Delta^{(2)}\cr
P^{D3}_{(-)} &=& \frac12 (1-\sigma_2\otimes \Gamma_{D3})\; ,
\end{eqnarray} 
where $\sigma_1,\sigma_2$ are the standard Pauli matrices described in appendix A.
Hence the final form for the fermionic part of the D3-brane actions is
\begin{eqnarray}
S^{(2)}_{D3}=iT_{D3}\int{d\xi^4\;e^{-\phi}\sqrt{-g}\;\bar{y}\;P^{D3}_{(-)}(\Gamma^I \breve D_I -\breve \Delta)y}\;.
\end{eqnarray}

One obtains the action for any other D-branes analogously. After a long but straightforward calculation, one finds the following general action for any Dp-brane\footnote{ Recall that we use the superspace convention for differential forms, i.e. $w^{(p)}=\hbox{${1\over p!}$}dx^{m_1}\wedge\cdots \wedge dx^{m_p} w_{m_p\cdots m_1}$, which arises naturally from the Hassan formalism.}:
\bea
S_{Dp}=-T_{Dp}\int{d^{p+1}\xi e^{-\phi}\sqrt{-\det(g+f)}}+T_{Dp}\int{\Sigma( \;C\;e^{-f})} \;+ \nn
+\;i T_{Dp}\int{ d^3\xi\;e^{-\phi}\sqrt{-g}\;\bar{y}\;P^{Dp}_{(-)}\left(\Gamma^i\breve{D}_i -\breve{\Delta}\right)y}\;,
\label{fdb}
\eea
where,
\bea
&& \hbox{for p odd:}\;\left\{
\begin{array}{cc}
\;\;\;\breve{D}_m=\hat{D}^{(0)}_m+\sigma_1\otimes\hat{W}_m &,\;\; \breve{\Delta}=\hat{\Delta}^{(1)}
+\sigma_1\otimes\hat{\Delta}^{(2)} \\ \\
y=\left(
\begin{array}{cc}
y_1\\
y_2\\
\end{array}\right), \\
\end{array}\right.
\eea
\bea
\hbox{while for p even:}\;\left\{
\begin{array}{cc}
&\breve{D}_m=D_m\;\;\;,\;\;\; \breve{\Delta}=\Delta \;\;\;\;\;\;\;\;\;\;\;\;\;\;\;\;\;\;\;\;\;\;\;\;\;\;\\\\
&y=\left(y_+\;+\;y_-\right) . \;.\;\;\;\;\;\;\;\;\;\;\;\;\;\;\;\;\;\;\;\;\;\;\;\;\;\;\\
\end{array}\right.
\eea
In addition, $P^{Dp}_{(-)}=\hbox{${1\over 2}$}(1-\breve{\Gamma}_{Dp})$, where $\breve{\Gamma}_{Dp}$
is given by,
\bea
\breve \Gamma_{D(4q)}&=&\hbox{${1\over (4q+1)!\sqrt{-g}}$}\epsilon^{i_1..i_{(4q+1)}}
\Gamma_{i_1..i_{(4q+1)}}\Gamma^{\ul{\varphi}}\;, \nn
\breve \Gamma_{D(4q+2)}&=&\hbox{${1\over (4q+3)!\sqrt{-g}}$}\epsilon^{i_1..i_{(4q+3)}}
\Gamma_{i_1..i_{(4q+3)}}\;, \nn
\breve \Gamma_{D(4q+1)}&=&-\sigma_1\otimes\hbox{${1\over (4q+2)!\sqrt{-g}}$}\epsilon^{i_1..i_{(4q+2)}}
\Gamma_{i_1..i_{(4q+2)}}\;, \nn
\breve \Gamma_{D(4q+3)}&=&-\sigma_2\otimes\hbox{${i\over (4q+4)!\sqrt{-g}}$}\epsilon^{i_1..i_{(4q+4)}}
\Gamma_{i_1..i_{(4q+4)}}\;, \nn
\eea
for $q=(0,1,2)$.
\vspace{1cm}

\section{Summary}

In this work we have expressed the Dp-brane actions in terms of component fields, neglecting interactions between fermions and the field $f$. Modulo these neglected terms, we have worked to second order in the fermions about an arbitrary bosonic supergravity background. Again modulo the neglected terms, these actions are k-symmetric by construction, though we have reserved the detailed relation to rigid supersymmetry for a future work \cite{mms}. We have used normal coordinate expansion techniques, which allowed us to write the full second order expansion in terms of covariant tensors. We have also obtained the T-duality rules for these actions using the formalism of Hassan \cite{has}.  
We emphasize that our results, i.e. (\ref{fdb}), yield the first explicit component-field presentation of these actions in generic bosonic backgrounds. We have not fixed k-symmetry, and all terms are covariant under bulk space-time transformations.  

Our methodology was as follows.  First, the M2-brane k-symmetric action was expanded using normal coordinates up to second order in the fermions following \cite{gk1}.  Next, we performed a single dimensional reduction to obtain the D2-brane action. Finally, by means of the T-duality map, we recovered the D$p$-brane actions for all other $p$.

A forthcoming paper \cite{mms} will compute the neglected interaction terms between gauge fields and fermions. A detailed discussion of the k-symmetries and supersymmetries will also be presented.  Such a treatment requires consideration of many details of gauge fixing schemes which we were able to avoid here.

\vspace{2cm}
\noindent
{\bf Acknowledgments}\\

We thank M. Grisaru, R. Myers and D. Zanon for useful discussions. L. Martucci and P. J. Silva were partially supported by INFN, MURST and by the European Commission RTN program HPRN-CT-2000-00131, in association with the University of Torino. D. Marolf and P. J. Silva were supported in part by NSF grant PHY00-98747 and by funds from Syracuse University. 


\section{Appendix A: Spinor conventions and Gamma matrix algebra}

This appendix is a list of spinor conventions.

In 11D superspace, we denote the general supercoordinates by $z^M$, where $M$ runs over the bosonic coordinates $x^m$, and fermionic coordinates $\theta^\mu$. Thus the curved index $M$ splits into $M=(m,\mu)$, where $m=0,1,...,10$, $\mu=1,2,...,32$. We use $A=(a,\alpha)$ for tangent space indices and we underline explicit tangent space indices (e.g., {\ul 0}, {\ul1}, etc.), to differentiate them from explicit space-time indices.

We take the metric to have signature $(-,+,...,+)$ and use the Clifford algebra \be
\{\Gamma^a,\Gamma^b\}=2\eta^{ab}\;,
\ee
where $\Gamma^a$ are real gamma matrices and $\eta^{ab}$ is the 11D Minkowski metric. 
We also set $\epsilon^{01\ldots}=1$ and use the notation 
$\Gamma_{a_1...a_n}=\Gamma_{[a_1}...\Gamma_{a_n]}$ 
denoting antisymmetrization with weight one; e.g. $\Gamma_{01} = \hbox{${1\over2}$}(\Gamma_0 \Gamma_1- \Gamma_1 \Gamma_0) = \Gamma_0 \Gamma_1$.

We use real Majorana anticommuting spinors of 32 components, denoted $y^\alpha$ or $\theta^\mu$. 
The conjugation operation is defined by,
\bea
\bar{y}=y^TC\;, \nn
\bar y_\beta = y^\alpha C_{\alpha\beta}\;,
\eea
where $T$ corresponds to transpose matrix multiplication; e.g. $y^\alpha C_{\alpha\beta}$ instead of $C_{\alpha\beta}y^\beta$, and $C_{\alpha\beta}$ is the antisymmetric charge conjugation matrix with inverse $C^{\alpha\beta}$. The indices of a spinor and a bispinor $M^\alpha_{\;\;\beta}$ are lowered and raised via matrix multiplication by $C$ so that we have
\bea
C_{\alpha\beta}C^{\beta\gamma}=\delta^{\;\;\gamma}_{\alpha}\;, \nn
M_\alpha^{\;\;\beta}=C_{\alpha\gamma}M^\gamma_{\;\;\delta}C^{\delta\beta}\;, \nn
\bar{\theta}M\xi=\bar\theta_\alpha M^\alpha_{\;\;\beta} \xi^\beta =\theta^\alpha M_{\alpha\beta} \xi^\beta\;.
\eea
We take $C=\Gamma^{\ul{0}}$. It should also be noted that for Majorana spinors 
like $y$, any expression $\bar{y}\Gamma_{a_1..a_n}y$ vanishes for $n=(1,2,5,6,9,10)$ but in general is non-vanishing for $n=(0,3,4,7,8)$. For Majorana-Weyl spinors, only the corresponding expressions for $n=(3,7)$ are non-vanishing.

Once one of the directions, say $x^{10}$, has been compactified and the corresponding 
$\Gamma^{\ul{10}}$ is identified with the chiral gamma matrix $\Gamma^{\ul{\varphi}}$, 
the fermionic coordinates appearing in 11D supergravity can be decomposed into two Majorana 
Weyl spinors (each of which we write in 32-component form). Thus in 
type IIA we may write
\be
y=y_++y_-\;\;\hbox{where}\;\;\Gamma^{\ul{\varphi}} y_\pm=\pm y_\pm\;.
\ee
In type IIB, we choose the two 32 real component chiral spinors $y_1,y_2$ to have positive 
chirality, and we write them together as a 64-component spinor of the form
\bea
y=\left(
\begin{array}{cc}
y_1\\
y_2\\
\end{array}\right). 
\eea
Taking the tensor products of the 32 $\times$ 32 component $\Gamma$ matrices
with the 2 $\times$ 2 identity operator yields the 64 $\times$ 64 matrices 
\bea
&&\Gamma^a = \left( \begin{array}{cc}
\Gamma^a & 0 \\
0  & \Gamma^a \\
\end{array}\right)\;. \nonumber
\eea
Finally, we use the usual Pauli matrices,
\[
\sigma^1 = \left( \ba{cc} 0 & 1 \\ 1 & 0 \ea \right)
\qquad \qquad \sigma^2 = \left( \ba{cc} 0 & -i \\ i & 0 \ea \right)
\qquad \qquad \sigma^3 = \left( \ba{cc} 1 & 0 \\ 0 & -1 \ea \right) \]


\section{Appendix B: supergravity}

This appendix is a list of supergravity conventions. We emphasizethat in all this paper we use the superspace convention for differential forms i.e.  
\[w^{(p)}=\hbox{${1\over p!}$}dx^{m_1}\wedge\cdots \wedge dx^{m_p} w_{m_p\cdots m_1}\] .

\subsection{11D supergravity}

Here we borrow some of the conventions and definitions directely from Grisaru and Knutt \cite{gk1}. We also use bold letters for pulled-back superfields in the main body of the paper. 

The theory is described in terms of the vielbein ${E}^A (x, \theta) = dZ^M {E_M}^A$ and three-form $A =(1/3!)E^CE^BE^A A_{ABC}$  satisfying torsion constraints and field-strength constraints respectively\cite{Howe, Cremmer}:
\bea
(a)~~~&&{T_{\a \b}}^c = -i (\Gamma^c)_{\a \b} \nonumber\\
(b)~~~&&{T_{\a \b}}^{\gamma} = {T_{\a b}}^c ={T_{a b}}^c=0\nonumber\\
(c)~~~&&H_{\a \b \gamma \d} = H_{\a \b \gamma d}=H_{\a bcd}=0 \nonumber\\
(d)~~~&&H_{\a \b cd}= i (\Gamma_{cd})_{\a \b}
\eea
with $H = dA =(1/4!)E^DE^CE^BE^AH_{ABCD}$ and
\be
H_{ABCD}= \sum_{(ABCD)}\nabla_A A_{BCD} +{T_{AB}}^E A_{ECD}.
\ee
These constraints put the theory on shell.
{}From the Bianchi identities $DT^A = E^B{R_B}^A$, $D{R_A}^B=0$ and
$dH=0$, or 
\bea
&&\sum_{(ABC)}({R_{ABC}}^D- \nabla_A{T_{BC}}^D-{T_{AB}}^E {T_{EC}}^D)=0
\nonumber\\
&&\sum_{(ABCD)}(\nabla_A{R_{BCD}}^E+{T_{AB}}^F{R_{FCD}}^E)=0
\\
&&\sum_{(ABCDE)}(\nabla_AH_{BCDE} +{T_{AB}}^F H_{FCDE})=0
\nonumber
\label{Bianchis}
\eea
one derives expressions for the remaining components of the torsion
and curvature.
\bea
\label{torsolutions}
(e)~~~&&{T_{a \b}} ^\gamma= \frac{1}{36}{(\d_a^b\Gamma^{cde}+\frac{1}{8}{\Gamma_a}^{bcde})_{\b}}^\gamma H_{bcde}
\nonumber\\
(f)~~~&&{T_{ab}}^\a=\frac{i}{42}{(\Gamma^{cd})}^{\a\b} \nabla_\b H_{abcd}\\
(g)~~~&&(\Gamma^{abc})_{\a\b}{T_{bc}}^ \b=0\nonumber
\eea
\bea
\label{cursolutions}
(h)~~~&&{R_{ab,\gamma}} ^\d= \nabla_a{T_{b \gamma}}^\d-\nabla_b{T_{a\gamma}}^\d +\nabla_{\gamma}
{T_{ab}}^\d+{T_{a \gamma}}^\e {T_{b \e}}^\d-{T_{b \gamma}}^\e {T_{a \e}}^ \d
\nonumber\\
(i)~~~&&R_{\a b,cd}=\frac{i}{2}[(\Gamma_{b})_{\a\b}{ T_{cd }}^\b -(\Gamma_c)_{\a\b}
 {T_{db}}^\b
+(\Gamma_d)_{\a\b}{ T_{cb}}^\b ]\\
(j)~~~&&R_{\a\b,ab}= -\frac{i}{6}\left[{(\Gamma^{cd})}_{\a \b} H_{abcd} + \frac{i}{24}
(\Gamma_{abcdef})_{\a \b}H^{cdef}\right] \nonumber
\eea
with
\be
{R_{AB \g}}^{\d} = \frac{1}{4}{R_{AB cd}}
{(\Gamma^{cd})_\g}^\d~~~.
\ee
 
We will need the following additional consequences of the Bianchi identities:
\bea
\label{addrel}
(k)~~~&&\nabla_\a H_{bcde}=- 6i(\Gamma_{[bc})_{\a\b} {T_{de]}}^\b \\
(l)~~~&&\nabla_\a R_{bc,de}=\nabla_bR_{\a c,de}-\nabla_cR_{\a b,de}
+{T_{b\a}}^{\gamma}R_{\gamma c,de}-{T_{c\a}}^{\gamma}R_{\gamma b,de} 
-{T_{bc}}^{\beta}R_{\b \a ,de} \nonumber
\eea
which can be used to relate higher components (in $\theta$) of field strengths 
and curvatures to lower components. 

We also note the three-form equation of motion, \be
\nabla^a H_{abcd} = - \frac{1}{1728}\varepsilon_{bcd e_1 \cdots e_8}
H^{e_1 \cdots e_4} H^{e_5 \cdots e_8}
\ee
This is a consequence of the
constraints.

Finally, we should state that all quantities in the expansion are evaluated 
at $z^M= (x^m,0)$ so that they involve only the $\theta=0$ components 
of the superfields and their derivatives. In particular we have 
(in Wess-Zumino gauge but in fact our expansion is completely supergravity 
gauge-covariant), with ${E_M}^A| = {E_M}^A(x,0)$
\bea
{E_m}^a|&=&{e_m}^a (x) \nonumber\\
{E_m}^\a |&=& \psi_m^\a(x)= 0 \nonumber\\
{E_\mu}^a|&=&0 \nonumber\\
{E_\mu}^\a |&=& \d_\mu^\a
\eea
as well as
\bea
{T_{cd}}^\a | = \hat{\psi}_{cd}^\a= 0,
\eea
the {\em supercovariantized} gravitino field strength. We also note that
the choice of Wess-Zumino gauge  implies that the  spinor covariant
derivative connection vanishes at $\theta=0$, (see, for example, {\em Superspace}, 
 eq. (5.6.8) \cite{Superspace})
\be
{\omega_{\a\b}}^\g | =0~~~~~~~~ .
\ee


\subsection{10D supergravity}

Here we use basically the same convention as in \cite{antoine,has}.

First, we note that two types of RR field strength appear in the literature of type II supergravity. In addition to $dC^{(n)}$, it is also useful to introduce the following field strength definitions:
\begin{eqnarray}
{\bf F}^{(1)} &=& dC^{(0)}\cr
{\bf F}^{(2)} &=& dC^{(1)} \cr
{\bf F}^{(3)} &=& dC^{(2)} -C^{(0)}\,H\cr
{\bf F}^{(4)} &=& dC^{(3)} -C^{(1)}\wedge H\cr
{\bf F}^{(5)} &=& dC^{(4)} -C^{(2)}\wedge H\ .
\end{eqnarray}

The type IIA bosonic part of the action is given by
\bea
 S_{IIA} &=&
\frac{1}{2\kappa_{10}^2}\int d^{10} x \sqrt{-g}
    \Big\{
    e^{-2\phi} \big[
    R +4\big( \partial{\phi} \big)^{2}
    -\frac{1}{2 \cdot3!} H^2\big] + \nn
   && - \frac{1}{2\cdot 2!} ({\bf F}^{(2)})^2 - \frac{1}{2\cdot 4!} ({\bf F}^{(4)})^2 \Big\}
+ \frac{1}{4\kappa_{10}^2}\int b\wedge dC^{(3)}\wedge dC^{(3)}
\eea
and the supersymmetry transformations for the gravitino $\psi_m$ and dilatino $\lambda$ are
\bea
\delta\psi_m &=& \left[\partial_m +\frac{1}{4} \omega_{mab}\Gamma^{ab}+\frac{1}{4\cdot 2!}H_{mab}\Gamma^{ab}\Gamma^{\ul{\varphi}} \; + \right. \nn
&&\left. + \frac18 e^\phi \big( \frac{1}{2!} {\bf F}^{(2)}_{ab}\Gamma^{ab}\Gamma_m\Gamma^{\ul{\varphi}}+ \frac{1}{4!}{\bf F}^{(4)}_{abcd}\Gamma^{abcd}\Gamma_m\big)\right]\epsilon \ , \nn
\delta\lambda &=& \left[ \frac12 \left( \Gamma^m \partial_m\phi + \frac{1}{2\cdot 3!}H_{abc}\Gamma^{abc}\Gamma^{\ul{\varphi}}\right) \; +\right.\nn
&&+ \left. \frac{1}{8} e^\phi \left( \frac{3}{2!} {\bf F}^{(2)}_{ab}\Gamma^{ab}\Gamma^{\ul{\varphi}}+ \frac{1}{4!} {\bf F}^{(4)}_{abcd}\Gamma^{abcd}\right)\right] \epsilon \ .
\end{eqnarray}

The type IIB bosonic part of the action is given by
\bea
S_{IIB}&& =
\frac{1}{2\kappa_{10}^2}\int d^{10} x \sqrt{-g}
    \Big\{
    e^{-2\phi} \big[
    R +4\big( \partial{\phi} \big)^{2}
    -\frac{1}{2 \cdot3!} H^2\big] + \nn
   && - \frac{1}{2} ({\bf F}^{(1)})^2 - \frac{1}{2\cdot 3!} ({\bf F}^{(3)})^2 
- \frac{1}{4\cdot 5!} ({\bf F}^{(5)})^2  \Big\}
+ \frac{1}{4\kappa_{10}^2}\int b\wedge dC^{(2)}\wedge dC^{(4)}
\eea
and the supersymmetry transformations for the gravitino $\psi_m$ and dilatino $\lambda$ are,
\bea
\delta\psi_{(1,2)m} &=& \left[\partial_m +\frac{1}{4} \omega_{mab}\Gamma^{ab}\pm\frac{1}{4\cdot 2!}H_{mab}\Gamma^{ab} \; + \right. \nn
&&\left .+ \frac18 e^\phi \left(\mp {\bf F}^{(1)}_a\Gamma^a - \frac{1}{3!} {\bf F}^{(3)}_{abc}\Gamma^{abc}\mp
\frac{1}{4!}{\bf F}_{abcd}\Gamma^{abcd}\right)\Gamma_m \right]\epsilon_{(1,2)}\;, \nn
\delta\lambda_{(1,2)} &=& \left[ \frac12 \left( \Gamma^m \partial_m\phi \pm\frac{1}{2\cdot 3!}H_{abc}\Gamma^{abc}\right)\;+ \right. \nn 
&&\hspace{2,6cm}\left. + \frac{1}{2} e^\phi \left( \pm  {\bf F}^{(1)}_{a}\Gamma^{a}+
\frac{1}{2\cdot 3!} {\bf F}^{(3)}_{abc}\Gamma^{abc}\right) \right]\epsilon_{(1,2)}\; .
\eea

In the above expressions $\kappa_{10}=2\pi[(2\pi l_s)^8g_s^2]^{-1}$ and for the type IIB case, we use the convention that the self duality constraint on ${\bf F}^{(5)}$ is imposed by hand at the level of the equations of motion.


\section{Appendix C: Useful formulas for pulled-back fields}

Here we collect some useful results on the T-duality transformation rules for the pulled-back fields.
As in the main text, we use $f \sim O(y)$.

\subsubsection{From $Dp$ to $D(p+1)$}

Applying the rules for the pulled-back fields introduced in section \ref{td}, remembering that $x^9 \leftrightarrow A_9$ under T-duality, and neglecting terms of order $O(y)$ we find,
\begin{eqnarray}
P_{Dp}[e^a_{(-)i}]&=&P_{D(p+1)}[e^a_{i}]-P_{D(p+1)}[g_{99}]^{-1}P_{D(p+1)}[g_{9i}]P_{D(p+1)}[e^a_{9}]+O(y)\cr
&&\cr
P_{Dp}[e^a_{(+)i}]&=&P_{D(p+1)}[e^a_{i}]-P_{D(p+1)}[g_{99}]^{-1}P_{D(p+1)}[g_{9i}]P_{D(p+1)}[e^a_{9}]+O(y)\cr
&&\cr
P_{Dp}[\tilde g_{ij}]&=&P_{D(p+1)}[g_{ij}] -P_{D(p+1)}[g_{i9}]P_{D(p+1)}[g_{99}]^{-1}P_{D(p+1)}[g_{9j}]+O(y)\cr
&&\cr
P_{Dp}[\tilde g^{ij}]&=&P_{D(p+1)}[g^{ij}] + O(y)\; .
\end{eqnarray}

\bigskip
For $p$ even (from IIA to IIB), we have
\begin{eqnarray}
P_{Dp}[D^{(0)}_{i}y_+]&=&P_{D(p+1)}[\hat D^{(0)}_{i}y_1]- P_{D(p+1)}[g_{99}]^{-1}P_{D(p+1)}[g_{9i}]P_{D(p+1)}[\hat D^{(0)}_9 y_1]+O(y)\cr
&&\cr
P_{Dp}[D^{(0)}_{i}y_-]&=& -\Omega\left( P_{D(p+1)}[\hat D^{(0)}_{i}y_2]-P_{D(p+1)}[g_{99}]^{-1}P_{D(p+1)}[g_{9i}]P_{D(p+1)}
[\hat D^{(0)}_9 y_2]\right)+O(y)\cr
&&\cr
P_{Dp}[W_{i}y_+]&=&-\Omega\left( P_{D(p+1)}[\hat W_{i}y_1]-P_{D(p+1)}[g_{99}]^{-1}P_{D(p+1)}[g_{9i}]P_{D(p+1)}
[\hat W_9 y_1]\right)+O(y)\cr
&&\cr
P_{Dp}[W_{i}y_-]&=&P_{D(p+1)}[\hat W_{i}y_2]-P_{D(p+1)}[g_{99}]^{-1}P_{D(p+1)}[g_{9i}]P_{D(p+1)}
[\hat W_9 y_2]+O(y)\ ,
\end{eqnarray}

\bigskip
while, for $p$ odd (from IIB to IIA), we have
\begin{eqnarray}
P_{Dp}[\hat D^{(0)}_{i}y_1]&=&P_{D(p+1)}[D^{(0)}_{i}y_+]-P_{D(p+1)}[g_{99}]^{-1}P_{D(p+1)}[g_{9i}]P_{D(p+1)}
[D^{(0)}_9 y_+]+O(y)\cr
&&\cr
P_{Dp}[\hat D^{(0)}_{i}y_2]&=&\Omega\left( P_{D(p+1)}[D^{(0)}_{i}y_-]-P_{D(p+1)}[g_{99}]^{-1}P_{D(p+1)}[g_{9i}]P_{D(p+1)}[D^{(0)}_9 y_-]\right)+O(y)\cr
&&\cr
P_{Dp}[\hat W_{i}y_1]&=&\Omega\left( P_{D(p+1)}[W_{i}y_+]-P_{D(p+1)}[g_{99}]^{-1}P_{D(p+1)}[g_{9i}]P_{D(p+1)}
[W_9 y_+]\right)+O(y)\cr
&&\cr
P_{Dp}[\hat W_{i}y_2]&=&P_{D(p+1)}[W_{i}y_-]-P_{D(p+1)}[g_{99}]^{-1}P_{D(p+1)}[g_{9i}]P_{D(p+1)}
[W_9 y_2]+O(y)\ .
\end{eqnarray}

\bigskip
Finally, we need the following formula

\begin{eqnarray}
&&\frac{\epsilon^{i_0\ldots i_p}}{(p+1)!\sqrt{-det(P_{Dp}[\tilde g_{ij}])}} P_{Dp}[\Gamma_{(-)i_0}]\cdots
P_{Dp}[\Gamma_{(-)i_p}]\Omega=\cr
&&=\frac{-\epsilon^{I_0\ldots I_{p+1}}}{(p+2)!\sqrt{-det(P_{D(p+1)}[g_{IJ}])}} P_{D(p+1)}[\Gamma_{I_0}]\cdots
P_{D(p+1)}[\Gamma_{I_{p+1}}]\Gamma^{\ul\varphi}\ .
\end{eqnarray}

\subsubsection{From $D(p+1)$ to $Dp$}
Let us define  $P_{Dp}[E_{9i}]=\partial_i x^m E_{9m}$. Then we have the following identities

\begin{eqnarray}
P_{D(p+1)}[e^a_{(-)i}]&=& P_{Dp}[e^a_{i}]-g_{99}^{-1}P_{Dp}[E_{9i}]e^a_9\cr
P_{D(p+1)}[e^a_{(-)9}]&=& g_{99}^{-1}e^a_9\cr
&&\cr
P_{D(p+1)}[e^a_{(+)i}]&=& P_{Dp}[e^a_{i}]-g_{99}^{-1}P_{Dp}[E_{i9}]e^a_9\cr
P_{D(p+1)}[e^a_{(-)9}]&=& -g_{99}^{-1}e^a_9\cr
&&\cr
P_{D(p+1)}[\tilde g_{ij}]&=& P_{Dp}[g_{ij}] -P_{Dp}[E_{i9}]g_{99}^{-1}P_{Dp}[E_{9j}]+O(y)\cr
P_{D(p+1)}[\tilde g_{i9}]&=& g_{99}^{-1}P_{Dp}[E_{i9}]+O(y)\cr
P_{D(p+1)}[\tilde g_{9i}]&=& -g_{99}^{-1}P_{Dp}[E_{9i}]+O(y)\cr
P_{D(p+1)}[\tilde g_{99}]&=& g_{99}^{-1}\cr
&&\cr
P_{D(p+1)}[\tilde g^{ij}]&=& P_{Dp}[g^{ij}]+O(y)\cr
P_{D(p+1)}[\tilde g^{i9}]&=&- P_{Dp}[g^{ij}]P_{Dp}[E_{j9}]+O(y)\cr
P_{D(p+1)}[\tilde g^{i9}]&=& P_{Dp}[E_{9j}]P_{Dp}[g^{ji}]+O(y)\cr
P_{D(p+1)}[\tilde g^{99}]&=& g_{99}-P_{Dp}[E_{9i}]P_{Dp}[g^{ij}]P_{Dp}[E_{j9}]+O(y)\ .
\end{eqnarray}

\bigskip
\noindent Note that the above equations imply $P_{Dp}[E_{i9}]=-P_{Dp}[E_{9i}]+O(y)$.

For $p$ odd (from IIA to IIB), we have

\begin{eqnarray}
P_{D(p+1)}[D^{(0)}_i y_+]&=& P_{Dp}[\hat D_i^{(0)}y_1]-g_{99}^{-1}P_{Dp}[E_{i9}]\hat D_9^{(0)}y_1\cr
P_{D(p+1)}[D_9^{(0)}y_+]&=& g_{99}^{-1}\hat D_9^{(0)}y_1\cr
&&\cr
P_{D(p+1)}[D^{(0)}_i y_-]&=&-\Omega\left( P_{Dp}[\hat D_i^{(0)}y_2]-g_{99}^{-1}P_{Dp}[E_{9i}]\hat D_9^{(0)}y_2\right)\cr
P_{D(p+1)}[D_9^{(0)}y_-]&=& -\Omega g_{99}^{-1}\hat D_9^{(0)}y_2\cr
&&\cr
P_{D(p+1)}[W_i y_+]&=& -\Omega\left( P_{Dp}[\hat W_i y_1]-g_{99}^{-1}P_{Dp}[E_{9i}]\hat W_9 y_1\right)\cr
P_{D(p+1)}[W_9 y_+]&=& -\Omega g_{99}^{-1}\hat W_9 y_1\cr
&&\cr
P_{D(p+1)}[W_i y_-]&=&  P_{Dp}[\hat W_i y_2]-g_{99}^{-1}P_{Dp}[E_{i9}]\hat W_9 y_2\cr
P_{D(p+1)}[W_9 y_-]&=& -g_{99}^{-1}\hat W_9 y_2\ .
\end{eqnarray}

\bigskip

For $p$ even (from IIB to IIA), we have

\begin{eqnarray}
P_{D(p+1)}[\hat D^{(0)}_i y_1]&=& P_{Dp}[D_i^{(0)}y_+]-g_{99}^{-1}P_{Dp}[E_{i9}] D_9^{(0)}y_+\cr
P_{D(p+1)}[\hat D_9^{(0)}y_1]&=& g_{99}^{-1} \hat D_9^{(0)} y_+\cr
&&\cr
P_{D(p+1)}[\hat D^{(0)}_i y_2]&=&\Omega\left( P_{Dp}[D_i^{(0)}y_-]-g_{99}^{-1}P_{Dp}[E_{9i}]D_9^{(0)}y_-\right)\cr
P_{D(p+1)}[\hat D_9^{(0)}y_2]&=& \Omega g_{99}^{-1} D_9^{(0)}y_-\cr
&&\cr
P_{D(p+1)}[\hat W_i y_1]&=& \Omega\left( P_{Dp}[W_i y_+]-g_{99}^{-1}P_{Dp}[E_{9i}] W_9 y_+\right)\cr
P_{D(p+1)}[\hat W_9 y_1]&=& \Omega g_{99}^{-1} W_9 y_+\cr
&&\cr
P_{D(p+1)}[\hat W_i y_2]&=&  P_{Dp}[W_i y_-]-g_{99}^{-1}P_{Dp}[E_{i9}] W_9 y_-\cr
P_{D(p+1)}[\hat W_9 y_2]&=& -g_{99}^{-1} W_9 y_-\ .
\end{eqnarray}

Finally,

\begin{eqnarray}
&&\frac{\epsilon^{I_0\ldots I_{p+1}}}{(p+2)!\sqrt{-det(P_{D(p+1)}[\tilde g_{IJ}])}} P_{D(p+1)}[\Gamma_{(-)I_0}]\cdots
P_{D(p+1)}[\Gamma_{I_{(-)p+1}}]\Omega=\cr
&&=\frac{-\epsilon^{i_0\ldots i_p}}{(p+1)!\sqrt{-det(P_{Dp}[g_{ij}])}} P_{Dp}[\Gamma_{i_0}]\cdots
P_{Dp}[\Gamma_{i_p}]\Gamma^{\ul\varphi}\ .
\end{eqnarray}


\end{document}